\documentclass[english,10pt,prl,preprintnumbers,floatfix,aps,nofootinbib,notitlepage,showpacs,twocolumn]{revtex4-1}
\usepackage[T1]{fontenc}
\usepackage[latin9]{inputenc}
\usepackage{color}
\usepackage{verbatim}
\usepackage{amsmath}
\usepackage{amssymb}
\usepackage{graphicx}
\usepackage{esint}

\makeatletter

\providecommand{\tabularnewline}{\\}

\usepackage{fullpage}

\usepackage{latexsym}\usepackage{epsfig}

\usepackage{endnotes}

\newcommand{\ba}{\begin{eqnarray}}
\newcommand{\ea}{\end{eqnarray}}
\newcommand{\be}{\begin{equation}}
\newcommand{\ee}{\end{equation}}

\AtBeginDocument{

}

\usepackage{babel}

\makeatother

\usepackage{babel}
\begin{document}

\title{Non-linear evolution of the BAO scale in alternative theories of gravity}

\author{Emilio Bellini}
\email{emilio.bellini@icc.ub.edu}
\affiliation{ICC, University of Barcelona, IEEC-UB, Martí i Franquès
1, E-08028 Barcelona, Spain,}

\author{Miguel Zumalacárregui}
\email{zumalacarregui@thphys.uni-heidelberg.de}
\affiliation{Institut für Theoretische Physik, Ruprecht-Karls-Universität
Heidelberg, Philosophenweg 16, 69120 Heidelberg, Germany,}
\affiliation{Nordic Institute for Theoretical Physics, Roslagstullsbacken
23, 10691 Stockholm, Sweden}


\begin{abstract}
The scale of Baryon Acoustic Oscillations (BAO) imprinted in the matter
power spectrum provides an almost-perfect standard ruler: it
only suffers sub-percent deviations from fixed comoving length due
to non-linear effects. We study the BAO shift in the large Horndeski
class of gravitational theories and compute its magnitude in momentum
space using second-order perturbation theory and
a peak-background split. The standard prediction is affected by the
modified linear growth, as well as by non-linear gravitational effects
that alter the mode-coupling kernel. For covariant Galileon models,
we find a 14-45\% enhancement of the BAO shift with respect
to standard gravity and a distinct time evolution depending on the
parameters. Despite the larger values, the shift remains well below
the forecasted precision of next-generation galaxy surveys. Models
that produce significant BAO shift would cause large redshift-space
distortions or affect the bispectrum considerably. Our computation
therefore validates the use of the BAO scale as a comoving standard
ruler for tests of general dark energy models.
\end{abstract}

\maketitle


\section{Motivation}


One of the most exciting promises of modern cosmology is the possibility
of testing fundamental physics using the largest scales available
to observation \cite{Amendola:2012ys}. Among other developments,
the signatures of Baryon Acoustic Oscillations (BAO) have provided
an invaluable test of models for cosmic acceleration through their
imprint in the Cosmic Microwave Background \cite{Ade:2015rim}
and the distribution of Large Scale Structure (LSS) \cite{Eisenstein:1997ik}
using either galaxies \cite{Cole:2005sx,Eisenstein:2005su,Blake:2011en,Beutler:2011hx,Anderson:2013zyy}, the Lyman-$\alpha$ forest or quasars \cite{Font-Ribera:2013wce,Delubac:2014aqe}
(see Ref. \cite{Bassett:2009mm,Weinberg:2012es,Aubourg:2014yra} for reviews). To
an excellent approximation, the BAO signal in the LSS provides a comoving
standard ruler that traces the expansion of the universe and probes
the onset of cosmic acceleration.


Non-linear corrections are known to introduce a small departure from
the perfect standard ruler behavior, systematically shifting the BAO
scale towards smaller values at low redshift. This effect has been
well studied for cold dark matter cosmologies with a cosmological
constant using perturbation theory \cite{Eisenstein:2006nj,Smith:2007gi,Crocce:2007dt,Matsubara:2007wj,Padmanabhan:2009yr,Noh:2009bb,McCullagh:2012qy,Sherwin:2012nh}
and simulations \cite{Seo:2009fp,Rasera:2013xfa,Prada:2014bra} (for earlier works see \cite{Bharadwaj:1996qm,Meiksin:1998ra}). The
result is that the BAO scale imprinted in the matter distribution shrinks
by approximately $0.3\%$ at redshift zero \cite{Seo:2009fp,Prada:2014bra}.
However, this value relies on the assumption that gravity is Newtonian
in the scales of interest.

Little attention has been devoted to the non-linear BAO evolution
in more general theories of gravity. Since the shift in the BAO scale
is comparable to the sub-percent level of precision expected by forthcoming galaxy surveys \cite{Font-Ribera:2013rwa}
that aim to test such theories, it will be necessary to understand
the effects of non-standard gravity on the BAO scale to correctly
interpret the data in the next generation of dark energy experiments.

\section{Peak-Background split computation of the BAO shift}


Sherwin and Zaldarriaga have explained the BAO shift in terms of the
effect of long modes on the short scale power spectrum \cite{Sherwin:2012nh}.
In their picture, large overdense regions undergo less overall expansion,
reducing the size of the physical BAO scale with respect to the average
(see also \cite{Zumalacarregui:2012pq}). This effect is not compensated
by underdense regions, because cosmic structures in overdense regions
undergo more growth and give a larger contribution to the power spectrum. 
Therefore, local differences in expansion and growth
lead to a net shortening of the comoving
BAO scale, causing a small departure from the standard ruler behavior.
Alternativelly, the shift of the BAO scale can also be understood
as arising from contributions to the power spectrum which are off-phase
with respect to the linear prediction \cite{Crocce:2007dt,Padmanabhan:2009yr}.


The BAO shift can be studied by comparing the non-linear power spectrum
to a rescaled version of the linear one \cite{Padmanabhan:2009yr}:
\begin{equation}
P(k) \approx P_{11}(k/\alpha)=P_{11}(k)-(\alpha-1)kP_{11}'(k)+\cdots\,,\label{eq:alpha_k_def}
\end{equation}
where the shift can be read from the coefficient of the second term
(more sophisticated templates are actually used to obtain the BAO
scale from data, but we will stick to this description for simplicity).
Here and below the power spectrum is defined as 
$\langle \delta(\vec k)\delta(\vec k^\prime)\rangle \equiv (2\pi)^3\delta_D(\vec k + \vec k^\prime)P(k)$ and $P_{nm}\propto \langle \delta_n \delta_m\rangle$, where we expand the density contrast as $\delta = \delta_1 + \delta_2 + \cdots$.

One can compare Eq. (\ref{eq:alpha_k_def}) with the prediction from standard perturbation theory
\begin{equation}\label{eq:pt_schematic}
P(k)=\left(P_{11}\cdots+P_{1n}\right)+\left(P_{22}\cdots+P_{mn}\right)\,.
\end{equation}
All $P_{1n}$ contributions are proportional to $P_{11}(k)$ and thus
do not contribute to the second term in Eq. (\ref{eq:alpha_k_def})
\cite{Padmanabhan:2009yr}. Only the mode-coupling terms (second parenthesis)
do contribute to the shift, with the first of such contributions given by
\begin{equation}
P_{22}(k)=\int\frac{d^{3}q}{(2\pi)^{3}}4\left[F_{2}(\vec{k}-\vec{q},\vec{q})\right]^{2}P_{11}(\vec{k}-\vec{q})P_{11}(\vec{q})\,.\label{eq:P_22}
\end{equation}
Here $F_{2}$ is the second-order symmetrized mode-coupling
kernel \cite{Bernardeau:2001qr}. The rest of the computation in the
peak-background split approximation proceeds by expanding in $k/q$,
integrating with a cutoff at $k_{BAO}$ and extracting the coefficient
of $kP^\prime_{11}(k)$ from the result (see \cite{Sherwin:2012nh}
for further details). The long modes with $q\ll k\sim k_{BAO}$ describe
the effect of the large fluctuations on the smaller scales.


The computation can be generalized to alternative theories of gravity
by noting that the structure of the kernel $F_{2}$ is preserved on
sub-horizon scales, but each term acquires a time-dependent coefficient
$C_{i}(t)$ 
\begin{equation}
F_{2}(\vec{p},\vec{q})=C_{0}+C_{1}\mu\left(\frac{p}{q}+\frac{q}{p}\right)+C_{2}\left(\mu^{2}-\frac{1}{3}\right)\,,\label{eq:F2}
\end{equation}
which reduces to the standard constant values, $C_{0}=17/21$,
$C_{1}=1/2$ and $C_{2}=2/7$, under matter domination in the case
of standard gravity (we drop the time-dependence for notation convenience).
Explicit computations in the sub-horizon, quasi-static limit of Horndeski
theories with non-relativistic matter determine that the modifications
to the kernel coefficients are not independent \cite{Takushima:2013foa,Bellini:2015wfa}
and satisfy 
\begin{equation}
C_{1}=\frac{1}{2},\quad C_{0}+\frac{2}{3}C_{2}=1\,.\label{eq:constraints_kernel}
\end{equation}

The BAO shift can be read by plugging the generalized kernel 
(\ref{eq:F2}), into the mode-coupling power spectrum 
(\ref{eq:P_22}), expanding to leading order in $q/k$, performing
the integration and comparing with Eq.~(\ref{eq:alpha_k_def}). This
generalizes the result of Ref. \cite{Sherwin:2012nh} to
\begin{equation}
\alpha-1=\frac{2}{5}\left(2C_{0}-\frac{1}{2}\right)\left\langle \delta_{L}^{2}\right\rangle \,,\label{eq:shift_SZ_generalized}
\end{equation}
where Eqs.~(\ref{eq:constraints_kernel}) have been used to write the
result in terms of the monopole $C_{0}$. In the above expression
the integration over the momentum leads to the long mode variance
\begin{equation}
\left\langle \delta_{L}^{2}\right\rangle \equiv\int_{0}^{k_{BAO}}\frac{dq}{(2\pi)^{3}}4\pi q^{2}P_{11}(q,t)\approx\sigma_{r_{s}}^{2}(t)\,,\label{eq:contrast}
\end{equation}
where we use a cut-off at BAO scale, well estimated
by the sound horizon at the drag epoch $k_{BAO}\sim1/r_{s}(z_{d})$
\cite{Thepsuriya:2014zda}. Following Ref. \cite{Sherwin:2012nh}, we
use the square of the variance of the density field on a sphere of radius $r_s(z_d)$ for the computation of the BAO shift: $\left\langle \delta_{\Lambda}^{2}\right\rangle \approx\sigma_{r_{s}}^{2}(t)$. 
This gives a slight underestimate with respect to the shift measured in simulations of standard cosmology, but we expect comparison among models to be accurate.

LSS surveys observe galaxies, which are known to be biased with respect to the underlying matter distribution. The effects of non-linear density-halo bias can be parameterized as
\begin{equation}
\delta_{h}(x)=b_{1}\delta(x)+\frac{1}{2}b_{2}\delta^{2}(x)+\cdots\,.\label{eq:bias_model}
\end{equation}
where the bias parameters $b_1,\,b_2$ relate the matter and the halo overdensity (the above expansion should hold on large scales). The halo-halo power spectrum generalizing Eq. (\ref{eq:pt_schematic}) reads
\begin{equation}\label{eq:halo_power}
P_h(k)=b_1^2\left( P_{11} + P_{22} \right) + b_1 b_2 P_{\delta^2_1\, \delta_2} +\cdots\,,
\end{equation}
where second-order terms that do not involve mode coupling have been omitted as they do not contribute to the BAO shift.
The first parenthesis contains the linear and mode-coupling matter power spectrum
renormalized by the linear bias. The second term mixes second order
corrections with non-linear bias, and affects the BAO shift. This term is given by
\begin{equation}
P_{\delta^2_1\,\delta_2}(k) =\int\frac{d^{3}q}{(2\pi)^{3}}2F_{2}(\vec k-\vec q,\vec q)P_{11}(\vec k-\vec q)P_{11}(q)
\,,\label{eq:bias_term}
\end{equation}
where the only difference with Eq. (\ref{eq:P_22}) is that the kernel appears linearly. As before, one can compute the leading order expansion in $q/k$ and compare with a rescaled version of the linear halo power spectrum (cf. Eq. (\ref{eq:alpha_k_def}) multipled by $b_1^2$). Identifying the $kP^\prime_{11}(k)$ term yields the contribution to the shift, which now reads
\begin{equation}\label{eq:shift_SZ_generalized_bias}
(\alpha-1)\big|_{h}=\left(\frac{4}{5}C_{0}-\frac{1}{5}+\frac{2}{3}\frac{b_{2}}{b_{1}}\right)\left\langle \delta_{L}^{2}\right\rangle \,.
\end{equation}
This result generalizes equation (\ref{eq:shift_SZ_generalized}).
Note that the relations (\ref{eq:constraints_kernel}) for the perturbation theory kernels make the bias contribution to the shift independent of the theory of gravity, i.e. of any departure in the value of $C_0$. For this reason we will not consider non-linear bias in the next section.

\section{BAO shift in alternative theories of gravity}


We focus our analysis on theories within the Horndeski Lagrangian
\cite{Horndeski1974}, which contains many examples of interest for
cosmology including Brans-Dicke, $f(R)$, Chameleons, Kinetic Gravity
Braiding and covariant Galileons. Horndeski's theory also
contains the characteristic interactions that appear in consistent
theories of massive gravity and higher dimensional theories, and it
is thus expected to effectively describe some of their distinctive
features \cite{Heisenberg:2014kea}.\footnote{We will not consider viable extensions of Horndeski's theory \cite{Zumalacarregui:2013pma,Gleyzes:2014dya},
nor full theories containing interacting gravitons \cite{deRham:2010kj,Hassan:2011zd,Hinterbichler:2012cn}.
See Refs. \cite{Amendola:1272934,Clifton:2011jh,Joyce:2014kja,Koyama:2015vza}
for reviews on the cosmology of alternative theories of gravity}


Although our analysis is general, for the sake of simplicity we will
present results for a covariant Galileon model \cite{Deffayet:2009wt}
(see also \cite{Nicolis:2008in,Deffayet:2009mn}). We fix the Galileon
Lagrangian parameters and the cosmological parameters
to the best-fit models obtained by Barreira \emph{et al.} \cite{Barreira:2014jha}
(without massive neutrinos), which have zero cosmological constant.
We noticed that the quintic galileon model we present
has a gradient instability in the tensor sector. However we decided
to include it in our analysis since it is a good fit for the data and has interesting
properties at second-order in perturbation theory. Indeed, as we shall
see it produces large modifications of the dark matter kernel, which
can be detected by studying the bispectrum with current surveys \cite{Gil-Marin:2014baa,Gil-Marin:2014sta}.

Simpler scalar-tensor theories (such as Brans-Dicke, Chameleons or
$f(R)$) lead to very constrained modifications and can not produce
large contributions to the Kernel \cite{Bellini:2015wfa}. For such
models the only sizeable contributions to the BAO shift stem from
the enhancement of linear growth, and would thus be in conflict with measurements
of LSS clustering. If one further demands that these theories are
screened in the Galaxy or the Solar system, the range of the scalar
force is too short to even affect cosmological scales in the linear
regime \cite{Wang:2012kj}. An exception is given by non-universal
couplings to matter: Coupled dark matter models can significantly
increase the shift of the BAO scale \cite{Cervantes:2012yf}.

\begin{table}
\begin{centering}
\begin{tabular}{|c|cc|cc|cc|}
\hline 
Model & \multicolumn{2}{c|}{$\sigma_{r_{s}}$ } & \multicolumn{2}{c|}{2$C_{0}$ } & \multicolumn{2}{c|}{$\alpha_{k}-1$ {[}\%{]}}\tabularnewline
\hline 
\hline 
$\Lambda$ & \multicolumn{2}{c|}{0.067} & \multicolumn{2}{c|}{1.62 } & \multicolumn{2}{c|}{0.20}\tabularnewline
\hline 
Cubic & 0.071 & (7\%)  & 1.61 &  (-0.4\%)  & 0.23 &  (14\%) \tabularnewline
\hline 
Quartic & 0.073 & (9\%) & 1.58  & (-2\%)  & 0.23  &  (15\%)\tabularnewline
\hline 
Quintic & 0.071 & (7\%)  & 1.92  & (19\%) & 0.29  & (45\%)\tabularnewline
\hline 
\end{tabular}
\par\end{centering}

\caption{Density contrast, mode coupling kernel monopole and BAO shift in the matter power spectrum for
reference model and selected Galileon models at redshift zero (cf.~Fig.~\ref{fig:time-dependence}).
Values in parenthesis indicate the relative deviation with respect
to a Cosmological Constant model. Tracer bias can be added using Eq. (\ref{eq:shift_SZ_generalized_bias}).
\label{tab:BAO-shift,-kernels}}
\end{table}

\begin{figure*}[!t]
\begin{centering}
\includegraphics[width=0.49\textwidth]{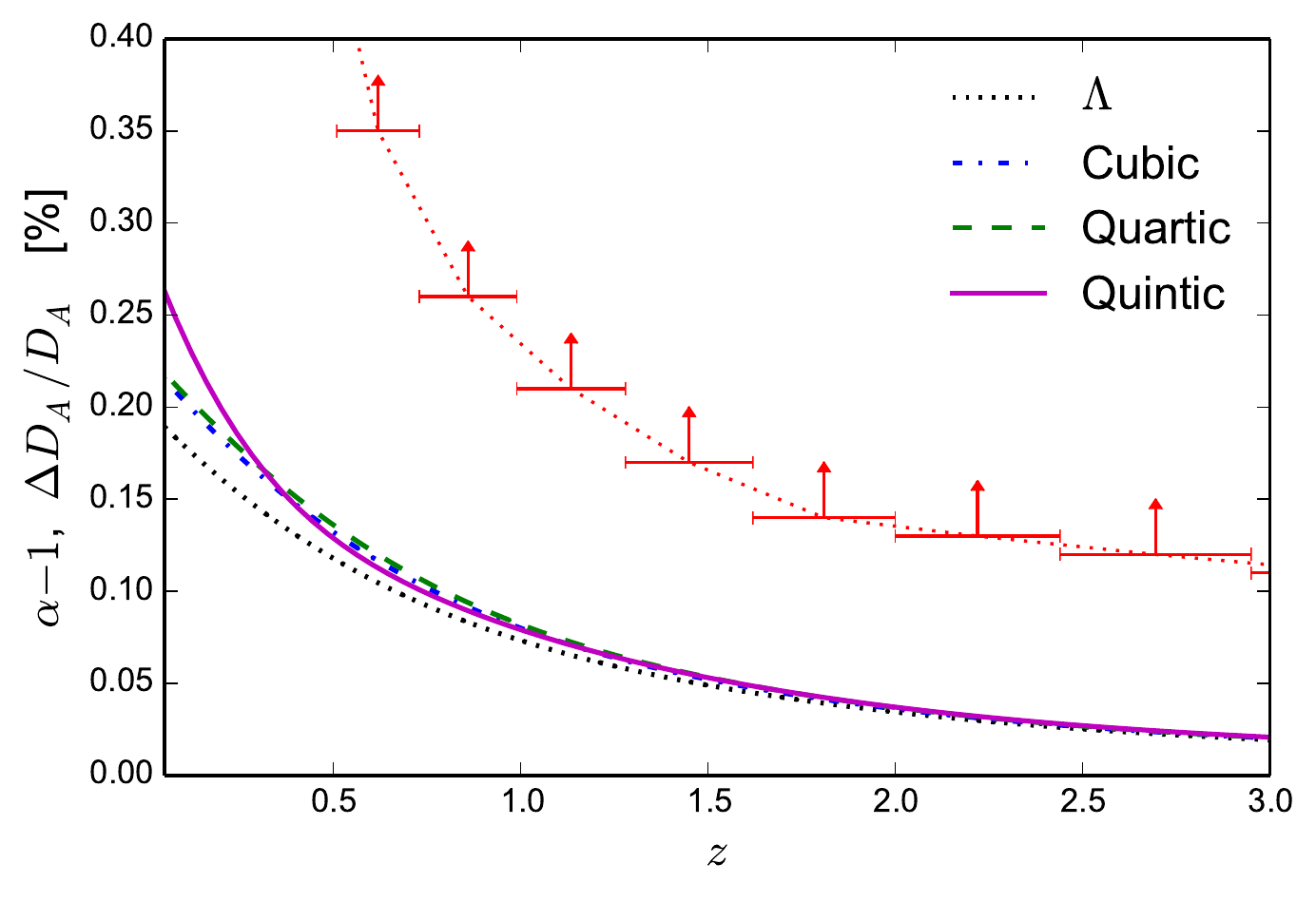}
\includegraphics[width=0.49\textwidth]{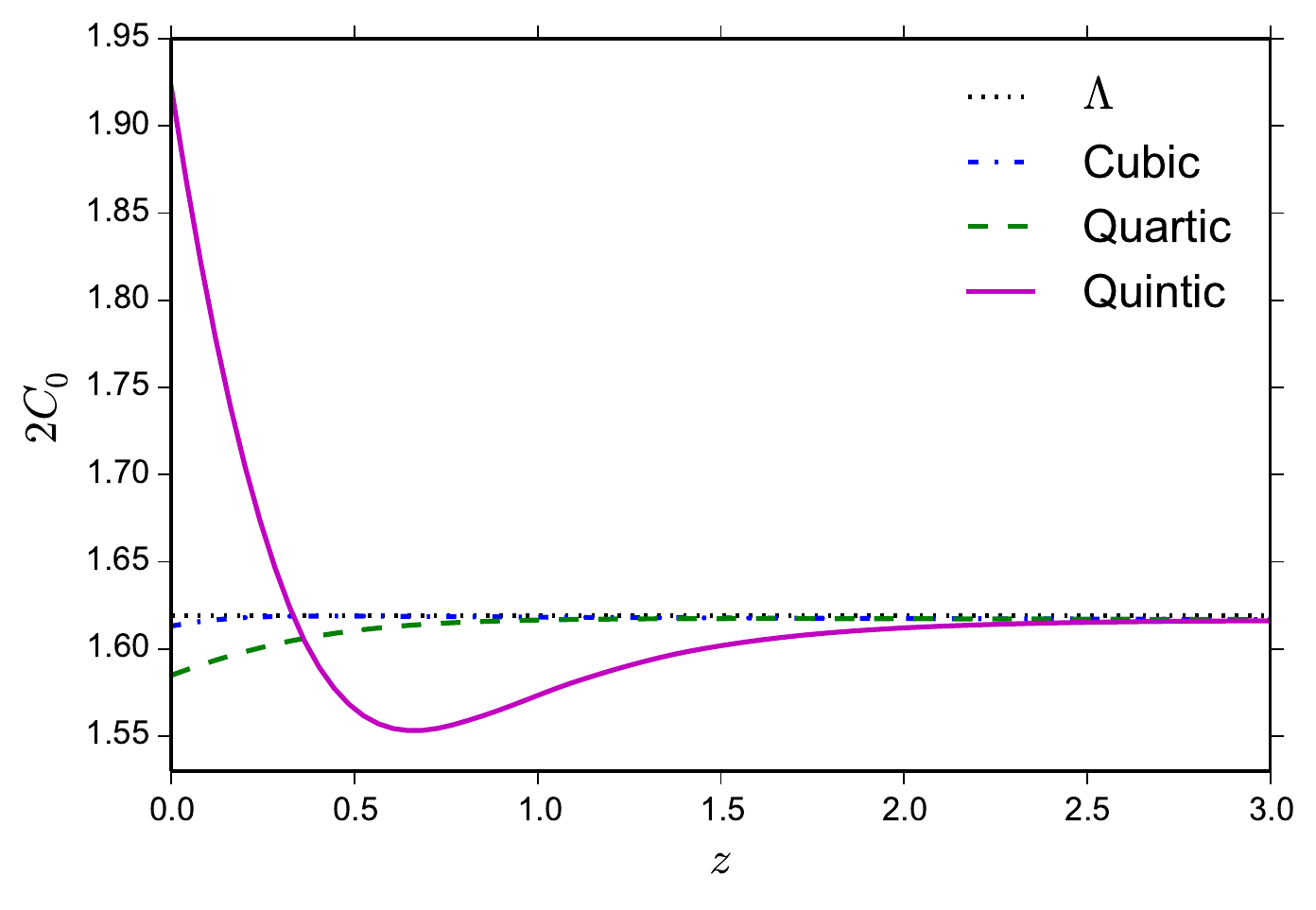}
\par\end{centering}
\caption{Time evolution of the BAO shift in the matter power spectrum (left panel) and integrated kernel (right panel) for standard and Galileon gravity models (cf.~Table
\ref{tab:BAO-shift,-kernels}). Red lines indicate expected sensitivity of $D_A$ across redshift bins from an optimistic BAO survey \cite{Weinberg:2012es} (see discussion around
Eq. (\ref{eq:BAO_uncertainty})).
\label{fig:time-dependence}}
\end{figure*}


The generalized Sherwin-Zaldarriaga formula (\ref{eq:shift_SZ_generalized})
depends on the theory of gravity in two ways: a correction from linear
physics, given by $\sigma_{r_{s}}^{2}$, and a modification of the
mode coupling kernel (\ref{eq:F2}), given by $C_{0}$.
We compute the evolution of the background, the linear
power spectrum and the density contrast $\sigma_{r_{s}}^{2}$ using
a modified version of the CLASS code \cite{Blas:2011rf,hi-class:2015}
based on the general description of Horndeski perturbations presented
in \cite{Bellini:2014fua} (see also \cite{Gleyzes:2013ooa}). The
computation of the non-linear corrections to the kernel follows the
approach of ref.~\cite{Bellini:2015wfa} (see also \cite{Bartolo:2013ws})
by taking the sub-horizon approximation and the quasi-static approximation (valid for covariant Galileons
on the scales of interest \cite{Sawicki:2015zya}). We will
also assume that the scale at which the model becomes strongly coupled
is smaller than the BAO scale. This is indeed the case for cubic and
quartic galileon models, as suggested by a comparison between fully
non-linear and linearized N-body simulations for Galileons \cite{Barreira:2013eea,Li:2013tda}.


The results for the BAO interesting quantities at
redshift zero are presented in Table \ref{tab:BAO-shift,-kernels}.
All the models considered tend to increase the density contrast $\sigma_{r_{s}}$
due to an enhanced effective force of gravity and the different background
expansion. The non-linear corrections to $C_{0}$ are highly dependent
on the model parameters, acquiring positive and negative sign and
ranging from sub-percent in the cubic, percent in the quartic, and
becoming fairly large in the quintic example.


The time evolution of the BAO shift and the mode-coupling corrections
are displayed in Figure \ref{fig:time-dependence}. Departures with
respect to standard gravity occur only at low redshift and become
largest in the accelerated era when the scalar field energy density
drives the cosmic expansion. Besides this general trend, each model
is characterized by a specific time dependence. Our results allow
to distinguish between a very soft non-linear regime in which the
mode coupling is mostly determined by interactions of the matter fluid
(standard gravity, cubic model) and large non-linear effects, as in
the case of the quintic model, with the quartic case being an intermediate
example. This is a consequence of the non-linear gravitational interactions
introduced in Horndeski's theory.

\section{Discussion}


Our results show an enhancement of the BAO shift with respect to the
standard prediction and provide the first estimate of this effect
for modified gravity. It is possible to compare the predicted shift
to the forecasted sensitivity of next-generation galaxy surveys. Let
us focus on measurements on the BAO scale transverse to the line of sight
$\theta_{BAO}=r_{s}/D_{A}$, where $D_{A}$ is the comoving
angular diameter distance (comparison with line-of-sight BAO yields
similar results). One can compare the two sources of uncertainty
\begin{equation}
\frac{\Delta D_{A}}{D_{A}}=\frac{D_{A}}{r_{s}}\Delta\theta+\frac{\Delta r_{s}}{r_{s}}\,,\label{eq:BAO_uncertainty}
\end{equation}
where the first term is the observational error (assuming known $r_{s}$)
and the second term is the systematic error induced by the shift,
$\Delta r_{s}/r_{s}=\alpha-1$. Weinberg \emph{et al.} have provided
an example forecast for an all-sky BAO survey in which the expected
error ranges from $2.8\%$ at $z=0.15$ to $0.1\%$ $z\gtrsim3.5$
(these data can be found in Table 2 of Ref. \cite{Weinberg:2012es}).\footnote{Their forecast also assumes density field reconstruction improvements
in the non-linear damping by a factor of 2. Since this procedure has
not been validated for general theories of gravity, we take the forecasted
precision as an optimistic bound.} Figure \ref{fig:time-dependence} compares both terms in Eq. (\ref{eq:BAO_uncertainty})
and shows that the BAO shift is well below the precision for all the
examples considered at any fiducial redshift. Note that the forecasted
precision is mainly limited by survey volume, implying that more sophisticated
observational setups will not be able to reduce the errors considerably.
More realistic forecasts based on specific surveys lead to
lower precision (see Ref. \cite{Font-Ribera:2013rwa}).

It is very unlikely that models more general than the ones considered
here can lead to sufficiently large shifts to bias the BAO scale measurements
while remaining compatible with other observations. The theoretical
prediction, Eq.~(\ref{eq:shift_SZ_generalized}),
allows one to identify two contributions to the shift: the modified
linear growth and the non-linear gravitational effects that modify
the mode-coupling kernel. Any theory of gravity with a very large
shift requires a considerable enhancement of at least one of these
contributions, which can be probed by observables other than BAO.%
\footnote{
Another possibility is that a modification of gravity enhances the BAO shift by producing a large value of the non-linear halo bias $b_2/b_1$. Such a possibility would however rely on the details of halo formation and its study would require methods other than cosmological perturbation theory.}

A large departure of $\sigma_{r_{s}}$ would be ruled out by redshift
space distortions or other clustering measurements. Similarly, large
corrections to the mode coupling kernel would induce large distortions
in the bispectrum (note that Eq.~(\ref{eq:constraints_kernel}) implies
that a $\gtrsim23.5\%$ increase in $C_{0}$ would change of sign
in the quadrupole term in $F_{2}$). We emphasize that these
non-linear gravitational effects are exclusive of fully fledged Horndeski
theories (cf. quartic and quintic example Galileons considered here)
and very suppressed in simpler scalar-tensor theories (e.g. Brans-Dicke,
$f(R)$) or cubic theories (e.g. our cubic example, Kinetic
Gravity Braiding \cite{Deffayet:2010qz} and limits of extra-dimensional
theories \cite{Dvali:2000hr}). Most works on higher order perturbation
theory for modified gravity have focused on the latter type of models
\cite{Scoccimarro:2009eu,Brax:2012sy,Bartolo:2013ws,Takushima:2013foa,Taruya:2014faa}.


Our findings validate the use of BAO measurements as a comoving standard
ruler for current and next-generation LSS surveys, at all redshifts
of interest and even for the most extreme theories of gravity. There
are several refinements that can improve our calculation,
such as including higher order perturbation theory corrections. Other
developments would be necessary in order to better connect these results
with observations, such as the inclusion of more sophisticated bias models (which
has been shown to affect the magnitude and time evolution \cite{Prada:2014bra})
and redshift-space distortions (which typically increase the magnitude
of the shift relative to real space). Finally, extending our result would allow to confirm the validity of density field reconstruction \cite{Padmanabhan:2008dd,White:2015eaa} of BAO for general theories of gravity.%
\footnote{The validity of BAO reconstruction schemes has been argued to rely exclusively in the equivalence principle \cite{Baldauf:2015xfa}. This assumption depends on the theory of gravity: it is valid for the Galileon models considered here but would be violated by Chameleon gravity \cite{Hui:2009kc}.}
Despite possible refinements, the smallness of the effects ensures the validity of our conclusions regarding the shift.


These are some initial steps in understanding the interplay between
extended theories of gravity and the BAO scale imprints on the distribution
of large scale structure. Further work should address other aspects
of LSS and BAO in general theories of gravity in order to optimize
the performance and model independence of the next-generation of dark
energy experiments. This will ultimately shed light on the optimal
strategy to test gravitational physics using future LSS surveys and
learn more on the connections between fundamental physics and cosmology.

\begin{acknowledgments}
\textbf{Acknowledgments:} We are very thankful to Diego Blas and Gerasimos
Rigopoulos for patiently sharing their knowledge about cosmological
perturbation theory, to Andreu Font-Ribera and Licia Verde for valuable comments on the manuscript, 
and to David Alonso, Bruce Bassett, Francisco
Castander, Antonio J. Cuesta, Juan Garcia-Bellido, Lam
Hui, Julien Lesgourgues, Nuala McCullagh, Massimo Pietroni, Paco Prada,
Alvise Raccanelli, Ignacy Sawicki, Marcel Schmittfull, Uros Seljak,
Blake Sherwin, Thomas Tram and Martin White for stimulating discussions
connected to this work. The work of EB is supported by the European
Research Council under the European Community's Seventh
Framework Programme FP7- IDEAS-Phys.LSS 240117. MZ is supported by
DFG through the grant TRR33 ``The Dark Universe''. We are both very
grateful to Nordita for hosting us in the program ``Extended theories
of gravity'', during which we completed an important part of the
work.

\end{acknowledgments}

\bibliographystyle{h-physrev}
\bibliography{bao_biblio}

\begin{thebibliography}{10}

\bibitem{Amendola:2012ys}
Euclid Theory Working Group, L.~Amendola {\em et~al.},
\newblock Living Rev.Rel. {\bf 16}, 6 (2013), 1206.1225.

\bibitem{Ade:2015rim}
Planck, P.~Ade {\em et~al.},
\newblock (2015), 1502.01590.

\bibitem{Eisenstein:1997ik}
D.~J. Eisenstein and W.~Hu,
\newblock Astrophys.J. {\bf 496}, 605 (1998), astro-ph/9709112.

\bibitem{Cole:2005sx}
2dFGRS, S.~Cole {\em et~al.},
\newblock Mon.Not.Roy.Astron.Soc. {\bf 362}, 505 (2005), astro-ph/0501174.

\bibitem{Eisenstein:2005su}
SDSS, D.~J. Eisenstein {\em et~al.},
\newblock Astrophys.J. {\bf 633}, 560 (2005), astro-ph/0501171.

\bibitem{Blake:2011en}
C.~Blake {\em et~al.},
\newblock Mon.Not.Roy.Astron.Soc. {\bf 418}, 1707 (2011).

\bibitem{Beutler:2011hx}
F.~Beutler {\em et~al.},
\newblock Mon.Not.Roy.Astron.Soc. {\bf 416}, 3017 (2011), 1106.3366.

\bibitem{Anderson:2013zyy}
BOSS, L.~Anderson {\em et~al.},
\newblock Mon.Not.Roy.Astron.Soc. {\bf 441}, 24 (2014), 1312.4877.

\bibitem{Font-Ribera:2013wce}
BOSS, A.~Font-Ribera {\em et~al.},
\newblock JCAP {\bf 1405}, 027 (2014), 1311.1767.

\bibitem{Delubac:2014aqe}
BOSS, T.~Delubac {\em et~al.},
\newblock Astron.Astrophys. {\bf 574}, A59 (2015), 1404.1801.

\bibitem{Bassett:2009mm}
B.~A. Bassett and R.~Hlozek,
\newblock (2009), 0910.5224.

\bibitem{Weinberg:2012es}
D.~H. Weinberg {\em et~al.},
\newblock (2012), 1201.2434.

\bibitem{Aubourg:2014yra}
E.~Aubourg {\em et~al.},
\newblock (2014), 1411.1074.

\bibitem{Eisenstein:2006nj}
D.~J. Eisenstein, H.-j. Seo, and .~White, Martin~J.,
\newblock Astrophys.J. {\bf 664}, 660 (2007), astro-ph/0604361.

\bibitem{Smith:2007gi}
R.~E. Smith, R.~Scoccimarro, and R.~K. Sheth,
\newblock Phys.Rev. {\bf D77}, 043525 (2008), astro-ph/0703620.

\bibitem{Crocce:2007dt}
M.~Crocce and R.~Scoccimarro,
\newblock Phys.Rev. {\bf D77}, 023533 (2008), 0704.2783.

\bibitem{Matsubara:2007wj}
T.~Matsubara,
\newblock Phys.Rev. {\bf D77}, 063530 (2008), 0711.2521.

\bibitem{Padmanabhan:2009yr}
N.~Padmanabhan and M.~White,
\newblock Phys.Rev. {\bf D80}, 063508 (2009), 0906.1198.

\bibitem{Noh:2009bb}
Y.~Noh, M.~White, and N.~Padmanabhan,
\newblock Phys.Rev. {\bf D80}, 123501 (2009), 0909.1802.

\bibitem{McCullagh:2012qy}
N.~McCullagh and A.~S. Szalay,
\newblock Astrophys.J. {\bf 752}, 21 (2012), 1202.1306.

\bibitem{Sherwin:2012nh}
B.~D. Sherwin and M.~Zaldarriaga,
\newblock Phys.Rev. {\bf D85}, 103523 (2012), 1202.3998.

\bibitem{Seo:2009fp}
H.-J. Seo {\em et~al.},
\newblock Astrophys.J. {\bf 720}, 1650 (2010), 0910.5005.

\bibitem{Rasera:2013xfa}
Y.~Rasera {\em et~al.},
\newblock Mon.Not.Roy.Astron.Soc. {\bf 440}, 1420 (2014), 1311.5662.

\bibitem{Prada:2014bra}
F.~Prada {\em et~al.},
\newblock (2014), 1410.4684.

\bibitem{Bharadwaj:1996qm}
S.~Bharadwaj,
\newblock Astrophys.J. {\bf 472}, 1 (1996), astro-ph/9606121.

\bibitem{Meiksin:1998ra}
A.~Meiksin, M.~J. White, and J.~Peacock,
\newblock Mon.Not.Roy.Astron.Soc. {\bf 304}, 851 (1999), astro-ph/9812214.

\bibitem{Font-Ribera:2013rwa}
A.~Font-Ribera {\em et~al.},
\newblock JCAP {\bf 1405}, 023 (2014), 1308.4164.

\bibitem{Zumalacarregui:2012pq}
M.~Zumalacarregui, J.~Garcia-Bellido, and P.~Ruiz-Lapuente,
\newblock JCAP {\bf 1210}, 009 (2012), 1201.2790.

\bibitem{Bernardeau:2001qr}
F.~Bernardeau, S.~Colombi, E.~Gaztanaga, and R.~Scoccimarro,
\newblock Phys.Rept. {\bf 367}, 1 (2002), astro-ph/0112551.

\bibitem{Takushima:2013foa}
Y.~Takushima, A.~Terukina, and K.~Yamamoto,
\newblock Phys.Rev. {\bf D89}, 104007 (2014), 1311.0281.

\bibitem{Bellini:2015wfa}
E.~Bellini, R.~Jimenez, and L.~Verde,
\newblock (2015), 1504.04341.

\bibitem{Thepsuriya:2014zda}
K.~Thepsuriya and A.~Lewis,
\newblock JCAP {\bf 1501}, 034 (2015), 1409.5066.

\bibitem{Horndeski1974}
G.~Horndeski,
\newblock International Journal of Theoretical Physics {\bf 10}, 363 (1974).

\bibitem{Heisenberg:2014kea}
L.~Heisenberg, R.~Kimura, and K.~Yamamoto,
\newblock Phys.Rev. {\bf D89}, 103008 (2014), 1403.2049.

\bibitem{Zumalacarregui:2013pma}
M.~Zumalacarregui and J.~Garcia-Bellido,
\newblock Phys.Rev. {\bf D89}, 064046 (2014), 1308.4685.

\bibitem{Gleyzes:2014dya}
J.~Gleyzes, D.~Langlois, F.~Piazza, and F.~Vernizzi,
\newblock (2014), 1404.6495.

\bibitem{deRham:2010kj}
C.~de~Rham, G.~Gabadadze, and A.~J. Tolley,
\newblock Phys.Rev.Lett. {\bf 106}, 231101 (2011), 1011.1232.

\bibitem{Hassan:2011zd}
S.~Hassan and R.~A. Rosen,
\newblock JHEP {\bf 1202}, 126 (2012), 1109.3515.

\bibitem{Hinterbichler:2012cn}
K.~Hinterbichler and R.~A. Rosen,
\newblock JHEP {\bf 1207}, 047 (2012), 1203.5783.

\bibitem{Amendola:1272934}
L.~Amendola and S.~Tsujikawa,
\newblock {\em {Dark energy: theory and observations}} (Cambridge Univ. Press,
  Cambridge, 2010).

\bibitem{Clifton:2011jh}
T.~Clifton, P.~G. Ferreira, A.~Padilla, and C.~Skordis,
\newblock (2011), 1106.2476.

\bibitem{Joyce:2014kja}
A.~Joyce, B.~Jain, J.~Khoury, and M.~Trodden,
\newblock Phys.Rept. {\bf 568}, 1 (2015), 1407.0059.

\bibitem{Koyama:2015vza}
K.~Koyama,
\newblock (2015), 1504.04623.

\bibitem{Deffayet:2009wt}
C.~Deffayet, G.~Esposito-Farese, and A.~Vikman,
\newblock Phys.Rev. {\bf D79}, 084003 (2009), 0901.1314.

\bibitem{Nicolis:2008in}
A.~Nicolis, R.~Rattazzi, and E.~Trincherini,
\newblock Phys.Rev. {\bf D79}, 064036 (2009), 0811.2197.

\bibitem{Deffayet:2009mn}
C.~Deffayet, S.~Deser, and G.~Esposito-Farese,
\newblock Phys.Rev. {\bf D80}, 064015 (2009), 0906.1967.

\bibitem{Barreira:2014jha}
A.~Barreira, B.~Li, C.~Baugh, and S.~Pascoli,
\newblock JCAP {\bf 1408}, 059 (2014), 1406.0485.

\bibitem{Gil-Marin:2014baa}
H.~Gil-Marin {\em et~al.},
\newblock (2014), 1408.0027.

\bibitem{Gil-Marin:2014sta}
H.~Gil-Marin {\em et~al.},
\newblock (2014), 1407.5668.

\bibitem{Wang:2012kj}
J.~Wang, L.~Hui, and J.~Khoury,
\newblock (2012), 1208.4612.

\bibitem{Cervantes:2012yf}
V.~D.~V. Cervantes, F.~Marulli, L.~Moscardini, M.~Baldi, and A.~Cimatti,
\newblock (2012), 1212.0853.

\bibitem{Blas:2011rf}
D.~Blas, J.~Lesgourgues, and T.~Tram,
\newblock JCAP {\bf 1107}, 034 (2011), 1104.2933.

\bibitem{hi-class:2015}
E.~Bellini, J.~Lesgourgues, and M.~Zumalacarregui,
\newblock in preparation.

\bibitem{Bellini:2014fua}
E.~Bellini and I.~Sawicki,
\newblock (2014), 1404.3713.

\bibitem{Gleyzes:2013ooa}
J.~Gleyzes, D.~Langlois, F.~Piazza, and F.~Vernizzi,
\newblock JCAP {\bf 1308}, 025 (2013), 1304.4840.

\bibitem{Bartolo:2013ws}
N.~Bartolo, E.~Bellini, D.~Bertacca, and S.~Matarrese,
\newblock JCAP {\bf 1303}, 034 (2013), 1301.4831.

\bibitem{Sawicki:2015zya}
I.~Sawicki and E.~Bellini,
\newblock (2015), 1503.06831.

\bibitem{Barreira:2013eea}
A.~Barreira, B.~Li, W.~A. Hellwing, C.~M. Baugh, and S.~Pascoli,
\newblock (2013), 1306.3219.

\bibitem{Li:2013tda}
B.~Li {\em et~al.},
\newblock JCAP {\bf 1311}, 012 (2013), 1308.3491.

\bibitem{Deffayet:2010qz}
C.~Deffayet, O.~Pujolas, I.~Sawicki, and A.~Vikman,
\newblock JCAP {\bf 1010}, 026 (2010), 1008.0048.

\bibitem{Dvali:2000hr}
G.~Dvali, G.~Gabadadze, and M.~Porrati,
\newblock Phys.Lett. {\bf B485}, 208 (2000), hep-th/0005016.

\bibitem{Scoccimarro:2009eu}
R.~Scoccimarro,
\newblock Phys.Rev. {\bf D80}, 104006 (2009), 0906.4545.

\bibitem{Brax:2012sy}
P.~Brax and P.~Valageas,
\newblock Phys.Rev. {\bf D86}, 063512 (2012), 1205.6583.

\bibitem{Taruya:2014faa}
A.~Taruya, T.~Nishimichi, F.~Bernardeau, T.~Hiramatsu, and K.~Koyama,
\newblock Phys.Rev. {\bf D90}, 123515 (2014), 1408.4232.

\bibitem{Padmanabhan:2008dd}
N.~Padmanabhan, M.~White, and J.~Cohn,
\newblock Phys.Rev. {\bf D79}, 063523 (2009), 0812.2905.

\bibitem{White:2015eaa}
M.~White,
\newblock (2015), 1504.03677.

\bibitem{Baldauf:2015xfa}
T.~Baldauf, M.~Mirbabayi, M.~Simonovic, and M.~Zaldarriaga,
\newblock (2015), 1504.04366.

\bibitem{Hui:2009kc}
L.~Hui, A.~Nicolis, and C.~Stubbs,
\newblock Phys. Rev. {\bf D80}, 104002 (2009), 0905.2966.

\end{thebibliography}

\end{document}